\documentclass[prb, twocolumn, showpacs, amsmath, amssymb,superscriptaddress]{revtex4-1}
\usepackage{amssymb,amsfonts,amsmath} 
\usepackage{graphicx}
\usepackage{color}
\usepackage{hyperref}
\usepackage{longtable}
\usepackage{bbm}
\usepackage{ulem}

\definecolor{darkred}{rgb}{0.75,0,0}
\begin{document} 

\title{Loschmidt-amplitude wave function spectroscopy and the physics of dynamically driven phase transitions}

\author{D.M.\ Kennes}
  \affiliation{Institut f\"ur Theorie der Statistischen Physik, RWTH Aachen University and JARA-Fundamentals of Future Information Technology, 52056 Aachen, Germany}
\affiliation{Max Planck Institute for the Structure and Dynamics of Matter, Center for Free Electron Laser Science, 22761 Hamburg, Germany}
\author{C.\ Karrasch}
\affiliation{Dahlem Center for Complex Quantum Systems and Fachbereich Physik, Freie Universit{\"a}t Berlin, 14195 Berlin, Germany}
\affiliation{Technische Universit\"at Braunschweig, Institut f\"ur Mathematische Physik, Mendelssohnstraße 3, 38106 Braunschweig, Germany}
\author{A.J.\ Millis}

\affiliation{Department of Physics, Columbia University, 538 West 120th Street, New Yok, NY 10027 USA}
\affiliation{Center for Computational Quantum Physics, Flatiron Institute, 162 5th Avenue, New York, NY 10010 USA}

\begin{abstract} 
We introduce the Loschmidt amplitude as a powerful tool to perform spectroscopy of generic many-body wave functions and use it to interrogate the wave function obtained after ramping the transverse field quantum Ising model through its quantum critical point. Previous results are confirmed and a more complete understanding of the population of defects and of the effects of magnon-magnon interaction or finite size corrections is obtained. The influence of quantum coherence is clarified.
\end{abstract}

\date{\today} 
\maketitle


Controlling quantum systems by application of tailored light pulses or other nonequilibrium protocols is established in quantum optics \cite{Streltsov17} and is rapidly growing in importance in condensed matter physics \cite{Mankowsky:Nonequilibrium,Basov:Towards,Tokura:Emergent}.  The goal is to design protocols that change the behavior of a system in some desired way, thereby accessing new regimes of behavior not accessible in equilibrium. Early applications were to few-body systems, but an important current issue is to extend the control to the  many-body regime.  A crucial issue, more severe in many-body systems than in few-body ones, is that when a system is exposed to an external perturbation, it will in general be excited out of its ground state, so that the distribution function as well as the Hamiltonian changes.  Typically, distribution function changes lead after relative short times to quasi-thermal states, which are often undesirable.

A tailored light field may be thought of as producing a time dependent Hamiltonian $H(t)$. As the time over which $H$ varies becomes longer and longer with respect to the basic energy scales of the system (such as the gap in the excitation spectrum), one expects that with increasing probability the system remains in the instantaneous ground state of the Hamiltonian at time $t$.  Many-body Hamiltonians often exhibit a dense spectrum of levels, making the applicability of these adiabatic-theorem ideas less obvious. One particularly interesting case is the `Kibble-Zurek' situation of a system tuned across a quantum critical point \cite{Kibble1976,Kibble1980,Zurek1985,Dutta2010,Dziarmaga2010,DelCampo2014}. The gap-closing at criticality means that the adiabatic theorem is necessarily violated, leading to creation of excitations.

These simple considerations highlight the need for theoretical methods of assessing the number and nature of the excitations created by a non-equilibrium drive. One way of characterizing a system is wave function spectroscopy: Given a wavefunction $\left|\Psi\right>$ at a specific time $t^\star$, one selects an eigenbasis  $\left|n\right\rangle$ (e.g., the eigenstates of $H(t=t^\star)$) and then constructs the corresponding density matrix $\hat{\rho}$ with elements $\rho_{nm}=\left\langle n \right |\left.\Psi\right\rangle \left\langle \Psi\right|\left. m \right\rangle $. For quantum many-body systems, however, $\rho_{nm}$ cannot be computed straightforwardly due to the prohibitively large size of the Hilbert space. 

In this letter, we present a powerful method for analyzing nonequilibrium wave functions which allows one to efficiently determine the spectral content of a given state without explicitly constructing eigenfunctions or finding eigenvalues. Our approach is based on the `Loschmidt amplitude' (its absolute value square is the Loschmidt echo \cite{Quan2010}), which is familiar from quantum optics \cite{Peres1984,Pastawski1995} and which has been employed in the pioneering work of Silva \cite{Silva2008} to characterize the work done by the application of a nonequilibrium perturbation \cite{Chenu2017} and by Pandey et. al. to analyze many-body localization \cite{spec1}. We apply the method to the one dimensional transverse field Ising model tuned through the order-disorder transition, uncovering  new  results that point to the importance of quantum coherence in Kibble-Zurek physics.  

{\it Methods---} Consider a closed system with a wave function $\left|\Psi(t)\right>$ obtained by the forward time evolution of some initial state $\left|\Psi_0\right\rangle$ with respect to a Hamiltonian $H(t)$.  Now choose a time $t^\star$ and form
\begin{equation}
L(t^\star,\omega)=\int_{-\infty}^{\infty} dt^\prime\left\langle\Psi(t^\star)\right|e^{i(H(t^\star)-E_0 - \omega)t'}\left|\Psi (t^\star\right)\rangle . \label{eq:loschmidt}
\end{equation}
Writing the integrand in terms of the eigenvalues $E_n$ and eigenstates $\left|n\right\rangle$ of $H(t^\star)$ gives the desired projection: $L(t^\star,\omega)\sim \sum_n |a_n|^2 \delta\left(\omega-(E_n-E_0)\right)$, where $a_n=\left\langle n\right|\left.\Psi(t^\star)\right\rangle$. 
If the time integral in Eq.~\eqref{eq:loschmidt} is taken over a finite but large $t_{\rm end}$, then the $\delta$-peaks are broadened (and one finds the usual Gibbs ringing) and contributions from energetically nearby eigenstates $\left|n\right\rangle$ with $E_n\approx E$ contribute to the same peak. In this sense, $t_{\rm end}$ restricts the frequency resolution of the wavefunction spectroscopy. If the spectrum is dense on the scale of this frequency resolution, then $L\sim |a(\omega)|^2 \rho(\omega)$, and we effectively sample the appropriate ``state density" $\rho(\omega)$ and corresponding ``occupation statistic" $|a(\omega)|^2$.

The key advantage of the Loschmidt amplitude is that the object $e^{i(H(t^\star)-E_0-\omega)t'}\left|\Psi\left(t^\star\right)\right\rangle$ 
can be calculated efficiently by several methods. Here, we employ the density matrix renormalization group (DMRG), which is an accurate numerical tool to study the equilibrium and non-equilibrium many-body physics of interacting one-dimensional systems. We use a real-time algorithm to directly evaluate Eq.~(\ref{eq:loschmidt}) (see Refs.~\onlinecite{White1992,Vidal2007a,Schollwock2011a,Kennes2016a} as well as \cite{SM}). For the model defined below in Eq.~(\ref{HTI}), we can reach time scales of $Jt_{\rm end}=120$ everywhere, which corresponds to a frequency resolution of $\Delta \omega/J \approx 2\pi/240\approx 0.026$.

\begin{figure}[t]
\centering
\includegraphics[width=0.8\columnwidth]{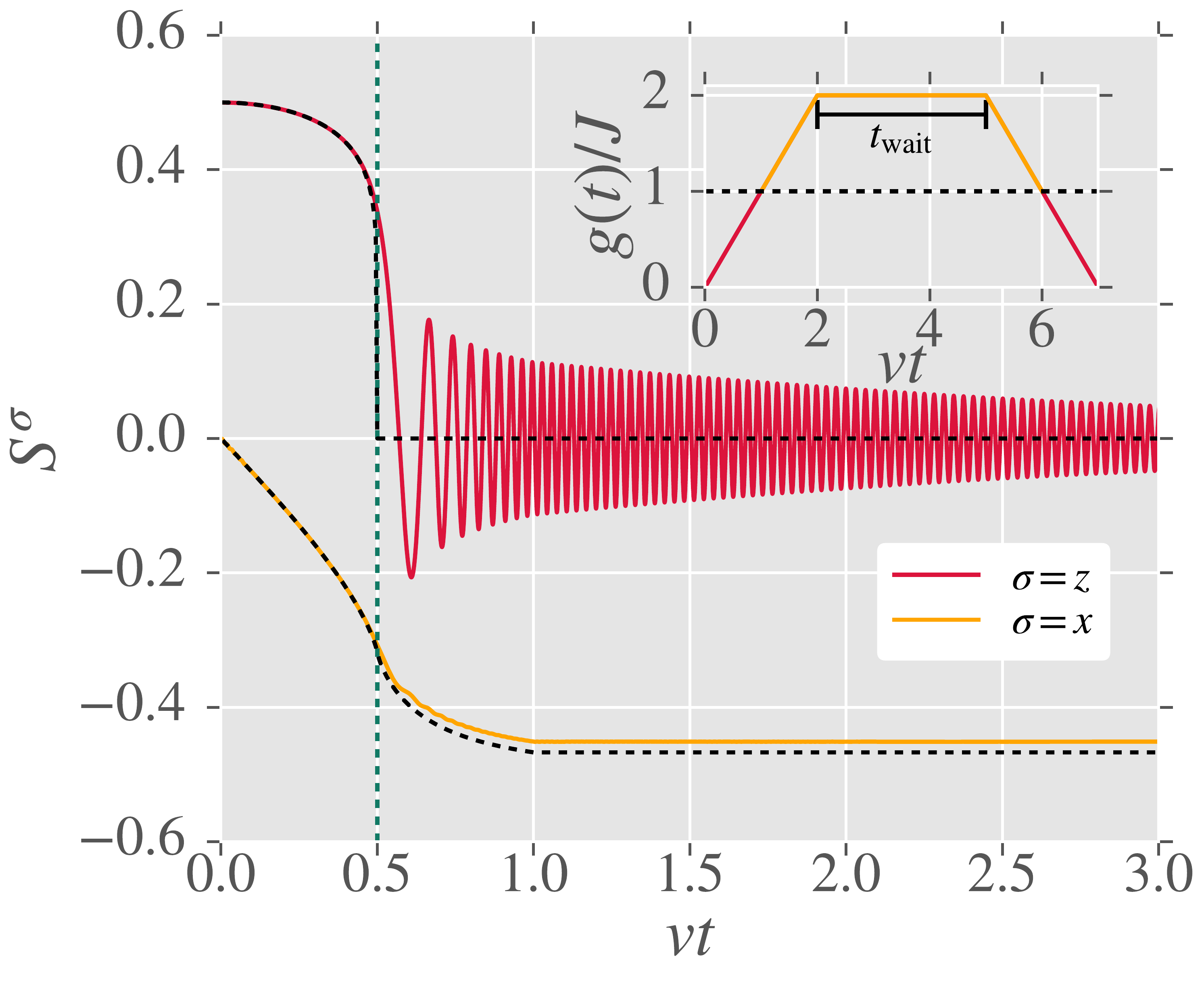}
\caption{$S^z$ and $S^x$ for a slow linear ramp through the QCP of the quantum Ising chain ($v/J=0.02$, $t_{\rm wait}J\to \infty$, $g_{\rm end}/J=2$, $N\to \infty$). The dashed green vertical line indicates the time when $g(t)=g_c=1$. The black dashed curves show the corresponding ground state expectation values with respect to the instantaneous Hamiltonian. Inset: Example of a temporal profile of $g(t)$ which governs the quench ($g_{\rm end}=2J$, $v/J=0.01$, $t_{\rm wait}=3/v$). The QCP at $g=g_c=1$ (dashed black horizontal line) separates the equilibrium ordered phase (red part of $g(t)$) from the disordered phase (orange part of $g(t)$). }
\label{fig:gt_S}
\end{figure}

{\it Transverse Field Ising Model---}
As an application, we consider the transverse field Ising model 
\begin{equation}
H(J,g)=\sum\limits_{i=1}^{N-1} -J \sigma^z_i\sigma ^z_{i+1}+\sum\limits_{i=1}^N g \sigma^x_i,
\label{HTI}
\end{equation}
where $\sigma^{x,y,z}_i$ denote Pauli matrices on site $i$. At $g=0$, the ground state of the model spontaneously breaks $\mathbb{Z}_2$ spin symmetry,  yielding  long ranged ferromagnetic order with moments aligned to the $\pm z$ direction and a gap to spin excitations.  For $0\leq g\leq J$, the model remains ordered, but the spins begin to cant into the $x$ direction and the $z$-component of the magnetization and the excitation gap concomitantly decrease. $g_c=J$ is a quantum critical point with gapless excitations, and for $g>J$ the system is again gapped, with the average spin pointing in the negative $x$ direction. 

We chose this model as it can solved exactly \cite{Lieb1961,Pfeuty1970,He2017}; the states are conveniently described \cite{Pfeuty1970} in a fermion representation, with the fermion creation operator related to the spin raising operator by a Jordan-Wigner string. At $g>J$ the states are classified into sectors labeled by number of fermions, $M$. The lowest energy state in a given sector has energy $2M(g-J)$, momentum $k=0$  and is the $M$ particle filled Fermi sea (note that for a finite system with $N$ sites and periodic boundary conditions the Jordan-Wigner factor shifts momenta so the allowed fermion momenta are odd integers times $\pi/N$ for even number of particles and even integers times $\pi/N$ for odd number of particles). Roughly, this ground state corresponds to exciting $M$ $k=0$ spin waves.  The higher energy states in the $M$ particle sector are a continuum of  particle-hole excitations above the $M$ particle Fermi sea and correspond roughly to exciting $k\neq 0$ spin waves

Here, we prepare the system in one of the two $g=0$ ferromagnetic ground states at time $t=0$ and for $t>0$ increases $g$ linearly with a `velocity' $v$ up to value $g_{\rm end}$ greater than unity, then hold $g$ at this value for a waiting time $t_{\rm wait}$, and eventually decrease it to zero. This `double ramp' is shown in the inset to Fig.~\ref{fig:gt_S}, and the equation for $g(t)$ is given in \cite{SM}. We also consider a `single ramp' in which $g$ is increased to a final value and then held indefinitely (by $t_{\rm wait}\rightarrow\infty$).


{\it Single Ramp through QCP---} Fig.~\ref{fig:gt_S} presents the average spin expectation values $S^{x,z}=\left\langle\sigma^{x,z}_{N/2}\right\rangle/2$ for a slow, single ramp through the QCP and $N\to \infty$. As $g$ is gradually increased away from $g=0$, at first $S^{x,z}$ remain indistinguishable from the ground state value calculated using the instantaneous Hamiltonian at time $t$, as expected from the adiabatic theorem \cite{Gell-Mann1951,Brouder2008}. However, when $g$ approaches the critical value $g=1$ (shown as a vertical dashed line), the adiabatic assumption breaks down; both $S^x$ and $S^z$ begin to deviate from their instantaneous values and for $g>1$ retain a ``footprint'' of the QCP-crossing. $S^x$ saturates to a constant value slightly smaller than the ground state one; the difference reflects the density of defects created as the system is tuned through the QCP and goes to zero as the ramp speed decreases \cite{adia1,adia2,adia3}. More intriguingly, $S^z$ exhibits coherent oscillations around the equilibrium value $S^z=0$ which only decay to zero at very long times. Coherent oscillations of the magnetization were not previously anticipated, 
 and we now turn to the Loschmidt methods to gain a better understanding. 

\begin{figure}[t]
\centering
\includegraphics[width=\linewidth]{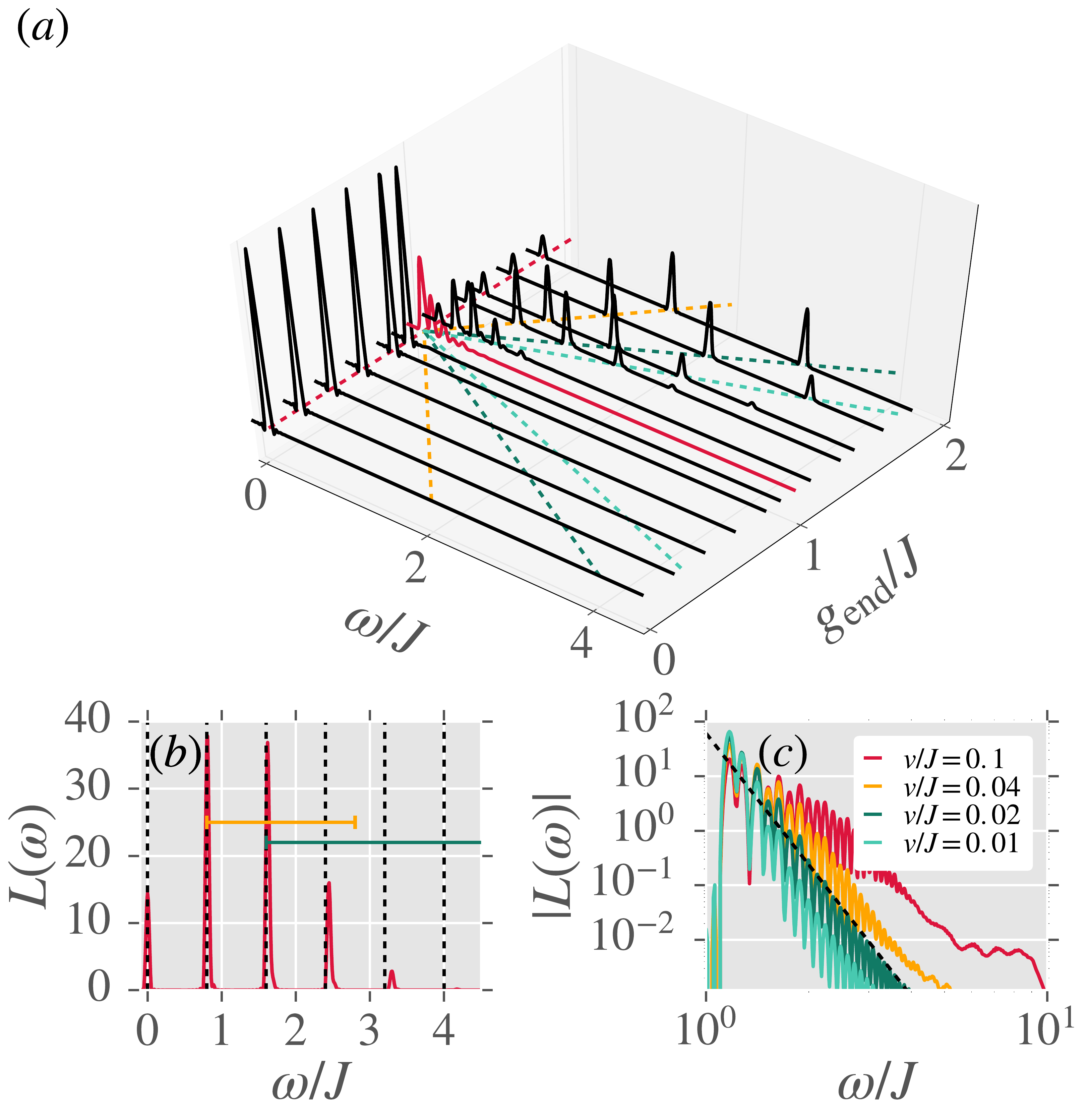}
\caption{ Wavefunctions spectrum $L(\omega)$ for a single forward ramp with $t_{\rm wait}J\to \infty$ and  $N=100$.  (a) $L(\omega)$ for fixed $v=0.02J$ and different $g_{\rm end}$. The evolution is adiabatic for $g_{\rm end}<g_c=1$; the spectrum shows a single peak corresponding to the ground state energy $\omega=0$ (dashed red line). After crossing the QCP, the wavefunction contains a superposition of eigenstates which are strongly localized in energy around integer multiples of the gap $\Delta= 2|g_\textnormal{end}-J|$ (colored dashed lines). The slice for $g_{\rm end}=g_c=J$ is highlighted in red. (b) $L(\omega)$ for fixed $v=0.02J$ and $g_{\rm end}/J=1.4$. Dashed vertical lines indicate $\omega= 2|g_\textnormal{end}-J|M$; yellow and green horizontal bars show the width of the first and second magnon band, respectively.  (c) $L(\omega)$ for a ramp that stops at the QCP $g_{\rm end}/J=1.0$ at different $v$. The dashed line shows a $\sim t^{-8}$ power-law.}
\label{fig:triple1}
\end{figure}

Panel (a) of  Fig.~\ref{fig:triple1} presents the Loschmidt amplitude $L(t,\omega)$ at different points during the evolution of $g$ from $g=0$ to $g=2J$,  against $\omega$ for different $t$ (parametrized here by the corresponding value of $g(t)$).  For times corresponding to $g(t)<J$, the evolution is adiabatic, and the probability for being in any state except the ground state associated with $g(t)$ is negligible. In the immediate vicinity of the critical point (slice in red), the spectrum is more complex, with a large number of low energy states excited. We note at the QCP $g/J=1$, the decay of $L(\omega)$ follows roughly a strongly decaying power-law $\sim \omega^{-8}$ if the ramp speed is slow (see Fig.~\ref{fig:triple1} (c)). For larger $g>1$, the situation simplifies again, and only a very small number of energies have a non-negligible contribution to the wave function. These states are at energies corresponding closely to integer multiples of the lowest, zero-momentum excitation energy $\Delta=2|g-J|$ (dashed colored lines) \cite{Lieb1961,Pfeuty1970,He2017} (for an analysis of the width of the peaks, which scales with $v$, see \cite{SM}). 

Panel (b) of Fig.~\ref{fig:triple1} examines in detail the Loschmidt signal at a time corresponding to $g=1.4J$. Only five states are present with any noticeable weight (the broadening is mainly due to the finite frequency resolution $\Delta \omega$; see \cite{SM}). The width of the first and second  magnon bands are shown as yellow and green bars, respectively; it is clear that the states appearing in the Loschmidt signal sit at or very near to  the lowest, zero-momentum energies of these bands; loosely, states with $M$ $k=0$ magnons.  The interpretation of the magnetization in terms of the wave function spectrum is deferred to \cite{SM}.

\begin{figure}[t]
\centering
\includegraphics[width=\columnwidth]{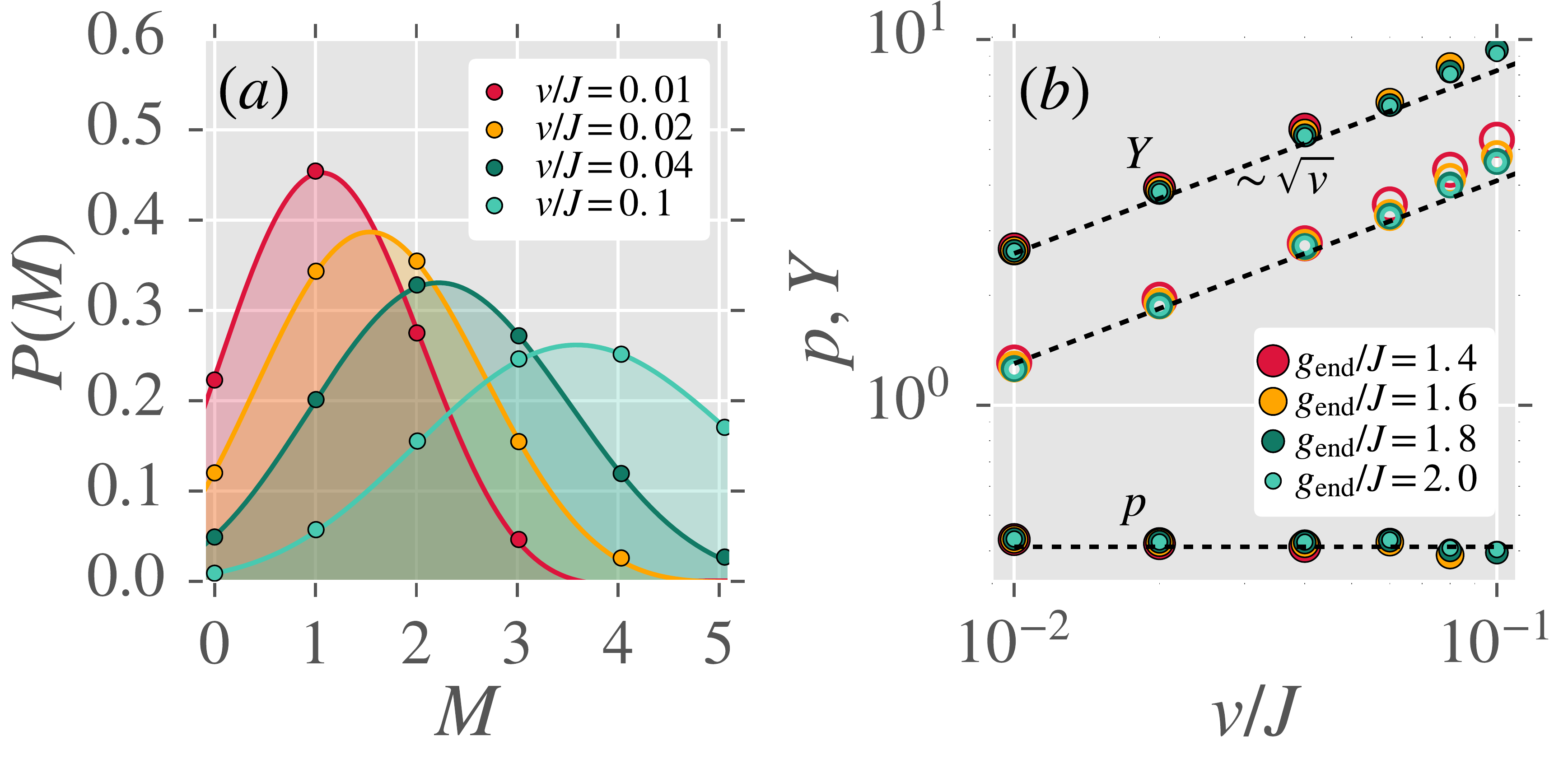}
\caption{(a) The distribution $P(M)$ of the number of states at excitation energy $2M(g-J)$ fitted to the  pseudo binomial distribution $P(M)=Y!/(Y-M)!/M! p^M(1-p)^{Y-M}$ ($g_{\rm end}/J=2$, $Jt_{\rm wait}\to \infty$, different speeds $v$). (b) Fitted values of $p$ and $Y$ as functions of ramp speed $v$ for different $g_{\rm end}$. Filled symbols are for $N=100$ and open ones for $N=50$ ($p$ for the latter not shown, which are on top of the $N=100$ data).  The data approximately collapses on single lines. Dashed lines show a power-law $Y\sim v^{1/2}$ (with a relative factor of 2 in the prefactor) as well as $p\sim C$. 
 }
\label{fig:triple3}
\end{figure}

We find that for $g_{\rm end}$ not too close to $1$, the wave function spectrum collapses if the frequency is plotted in units of $2|g_{\rm end}-J|$ (not shown). It is then useful to define the excitation probability in the  bottom $\omega_M$ of the $M$-th magnon band by $P(M)=\frac{1}{2\pi}\int_{\omega_M-2\Delta\omega}^{\omega_M+2\Delta\omega}L(\omega)d\omega$. Panel (a) of Fig.~\ref{fig:triple3} shows $P(M)$ for different speeds and demonstrates that the distribution can be fitted perfectly by a binomial form $P(M)=Y!/(Y-M)!/M! p^M(1-p)^{Y-M}$ with, crucially, a non-integer $Y$. The defect creation probability $p$ is found to be almost independent of ramp speed  $v$ and system size $N$. $Y$, which gives the mean number of defects created, scales as the square root of the ramp speed and is linearly proportional to system size, consistent with a constant defect creation density (see Fig.~\ref{fig:triple3} (b)). The $Y\sim \sqrt{v}$ relation is consistent with Kibble-Zurek scaling \cite{Kibble1976,Kibble1980,Zurek1985,Dutta2010,Dziarmaga2010}, which predicts $pY\sim v^\frac{d\nu}{1+\nu z}$, with known $d=z=\nu=1$.
From the exact solution in Ref.~\onlinecite{Dziarmaga05}, albeit for a different initial state, the average number of defects created during the ramp can be compared. We find that also the prefactor of $pY$ agrees within the 1\% regime with the anaytic prediction of $1/2\pi$.\cite{Dziarmaga05}
The ramp creates quantum defects, each with independent  probability $p$ per unit time and per unit length, with the dependence on the ramp velocity only via the expected value of the number of defects $pY$. We show that with the Loschmidt-amplitude spectroscopy we can extract the distribution of defects and are not restricted to models that are exactly solvable (as chosen in this work for benchmarking purposes).  

\begin{figure}[t]
\centering
\includegraphics[width=\columnwidth]{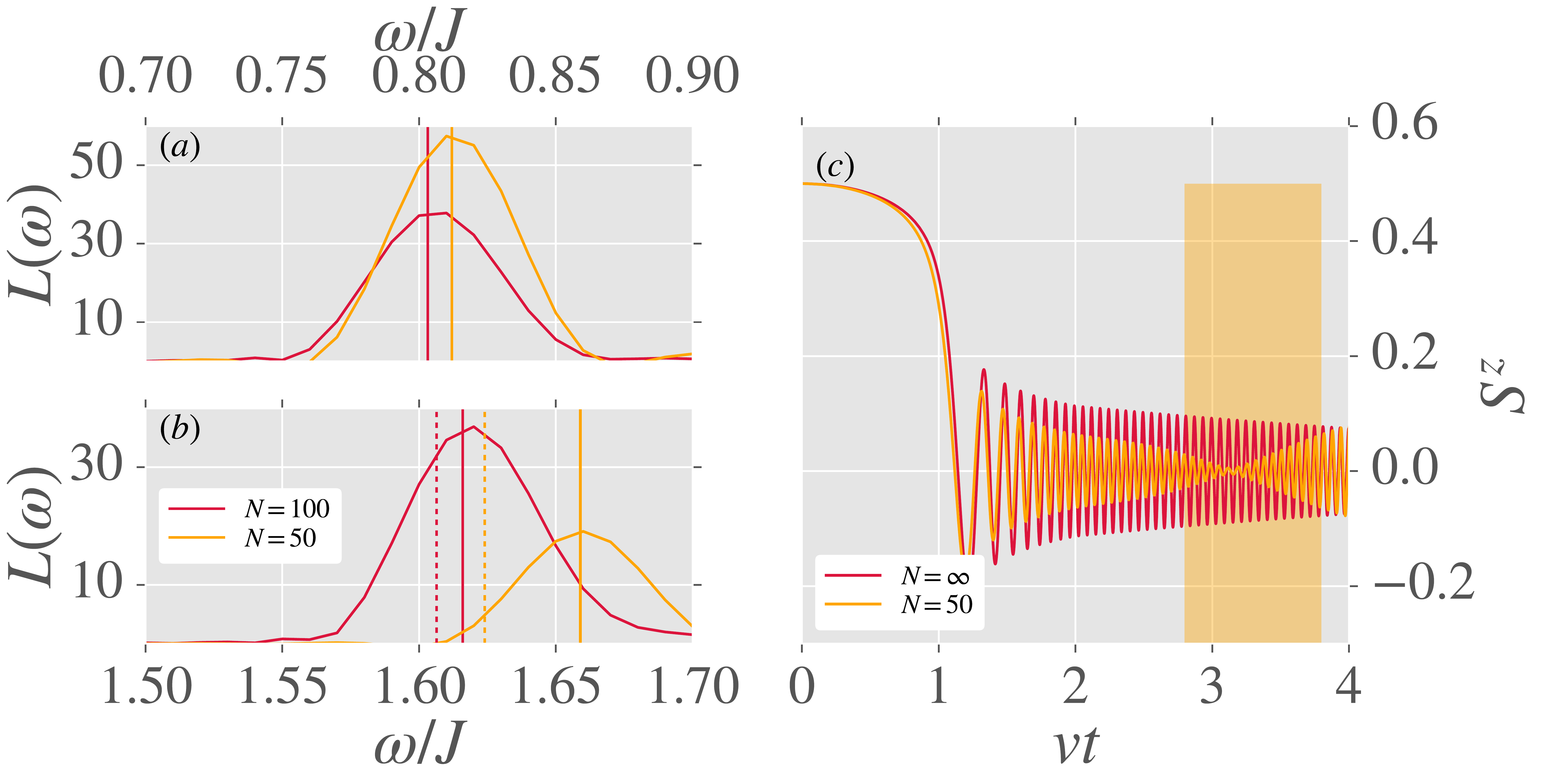}
\caption{Finite size dependence of the wavefunction spectrum: Zoom-in of (a) the one magnon as well as (b) the two magnon energy regime. Solid vertical lines shows the exact position of the lowest one or two magnon states. Dashed vertical lines in (b) indicate twice the one-magnon line of (a). (c) Time evolution of the spin expectation value $S^z$. Wavefunction spectroscopy allows for an analytic estimate for the beat frequency (see \cite{SM}) and for $N=50$ from (b) predicts that the first knot of the quantum beat should lies in the yellow shaded region (in agreement with the numerical data). The other parameters are as in Fig.~\ref{fig:triple1} (b).}
\label{fig:finN}
\end{figure}

It is important that Fig.~\ref{fig:gt_S} was obtained in the thermodynamic limit $N\to \infty$, while the wave function spectroscopy can only be performed on finite (yet large) systems. 
The result that the lowest energy state of each sector is $2M|g-J|$ holds only in the infinite system size limit.  In finite systems, these excitations are not at perfect integer multiples of each other \cite{Pfeuty1970}. In Fig.~\ref{fig:finN} (a) and (b), we show a zoom-in of the first two peaks of the wave function spectrum for two different system sizes as well as the exact excitation energies (vertical solid lines) \cite{Lieb1961,Pfeuty1970,He2017}. For comparison, dashed lines in (b) show twice the value of the solid lines in (a). Observables such as $S^z$, which connect eigenfunctions whose magnon numbers $M$ differ by one, thus display a superposition of oscillations at frequencies which are determined by the difference of the energies of the lower edges of two consecutive magnon-bands after the ramp. These frequencies are very close, so one finds a beating signal in $S^z$.
The beat frequency can be estimated analytically (see \cite{SM}); the approximated time interval (yellow shaded area in Fig.~\ref{fig:finN}) agrees well with the numerical data. Note that in more generic models, frequency shifts and quantum beats can also be induced by magnon-magnon interactions (in this sense we take finite $N$ as a proxy to more general magnon-magnon interactions). This highlights the importance of coherence on the dynamics of observables after a sweep through a QCP at finite system size and/or finite magnon-magnon interaction \cite{Kennes2018}. These results are of direct experimental relevance as current experiments on quantum simulations are performed at sizes of $N\sim 20-60$ \cite{Bernien2017}. 
%

\begin{figure}[t]
\centering
\includegraphics[width=\linewidth]{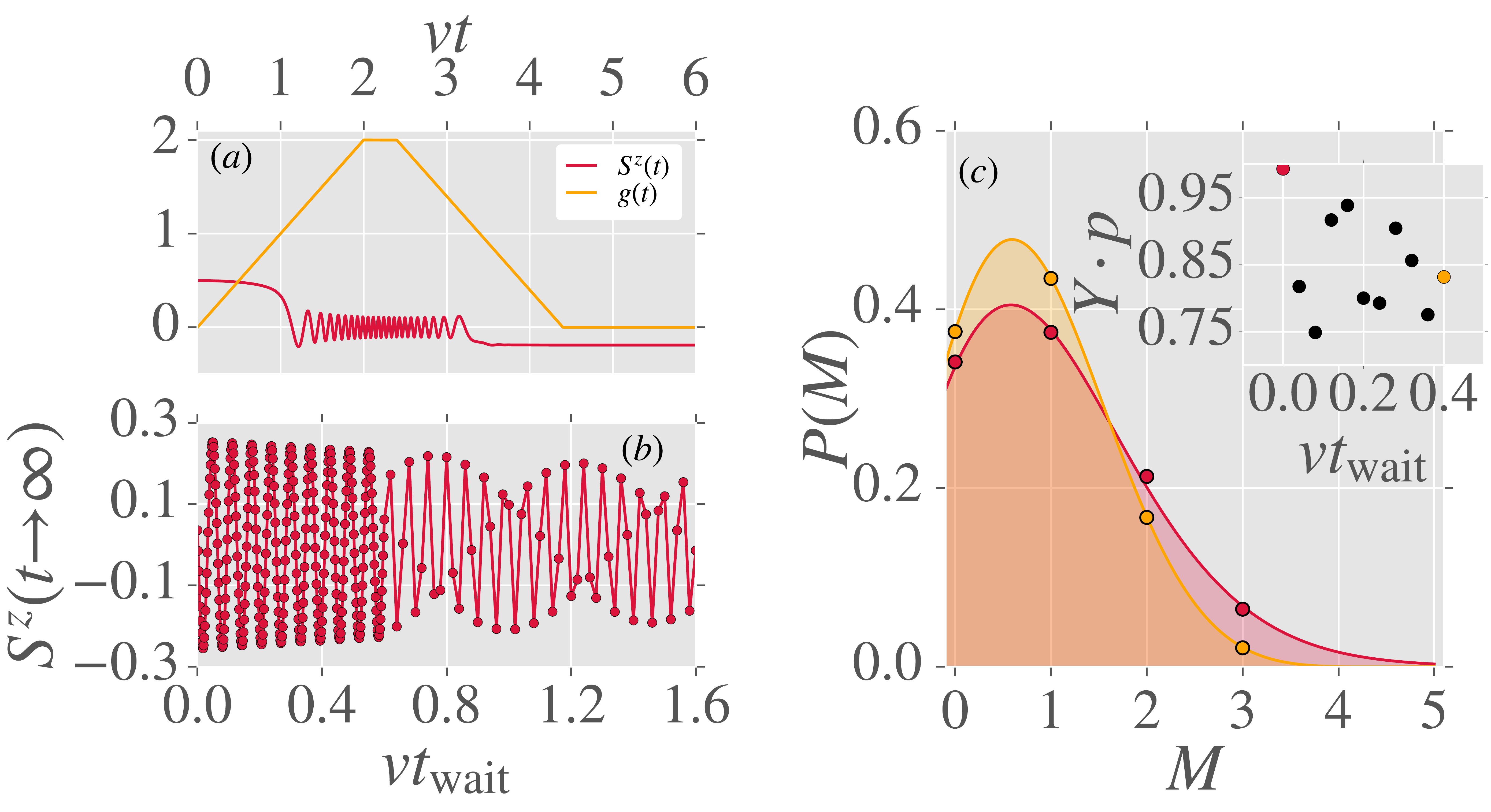}
\caption{Results for the double ramp. (a) Dynamics of $S^z(t)$ for a finite  $vt_{\rm wait}=0.4$ (and $v=0.02 J$, $g_{\rm end}=2J$, $N\to\infty$). After the second ramp is complete, $S^z(t)$ reaches a steady-state value. (b) The asymptotic value of $S^z(t\to\infty)$ for different waiting times $t_{\rm wait}$. The other parameters are as in (a). 
(c) Wavefunction spectrum $L(\omega)$ after the second ramp for two $vt_{\rm wait}=0.0$ (red) and  $vt_{\rm wait}=0.4$ (yellow). The other parameters are the same as in (a), but choosing $N=100$. Shaded regions show fits to Binomial distributions as in Fig.~\ref{fig:triple3}. The inset shows the fitted $Y\cdot p$ in dependency of the waiting time (including more $t_{\rm wait}$ then shown in the main panel), which is the average number of defects. }
\label{fig:triple2}
\end{figure}

{\it Ramp through QCP and back---} Finally, we consider the case of two slow ramps, one forward through the QCP as described above and then one backwards, with a finite waiting time $t_{\rm wait}$ in between (see Fig.~\ref{fig:triple2}(a)). 
The magnetization $S^z(t)$ shows long-lived oscillations after the first ramp but becomes time independent for $t>t_\textnormal{end}$  (see Fig.~\ref{fig:triple2} (a)), which one can easily understand from the fact that $\sigma^z_i$ commutes with $H$ for $g=0$. In Fig.~\ref{fig:triple2} (b), we demonstrate how the value of the magnetization for $t>t_\textnormal{end}$ can be tuned by adjusting the waiting time. The asymptotic value mimics the oscillations and decay of $S^z$ obtained during the waiting time, and the residual magnetization can hence be frozen in by the second ramp. Even after the second ramp is complete, the defects frozen into the final state approximately follow a binomial distribution (Fig.~\ref{fig:triple2} (c)). The fit parameters $p$ and $Y$ -- and thus the defect distribution -- can be controlled by tuning the waiting time. This implies that the phase differences in the pure wave function after the first ramp play an important role for the dynamics of crossing the QCP a second time. 
The fast dependence of $Y\cdot p$ on $t_{\rm wait}$ (inset to Fig.~\ref{fig:triple2} (c)) agrees with the  main oscillation $2|g_{\rm end}-J|$ found during the time when $g>1$.

{\it Summary and Outlook--- }We showed that quantum coherence can have prominent consequences for the dynamics encountered after a ramp through a QCP. Most notably, these consequences can manifest in a beating signature of the post-ramp dynamics of a given observable, which can be analyzed by virtue of the  wave function spectroscopy we introduced. This tool allows one to interrogate the many-body spectrum of large interacting systems and thus, e.g., characterize magnon-magnon interactions. We also illustrated how coherence in the quantum dynamics affects observables after a second ramp through the same QCP.

The established wave function spectroscopy could yield valuable insights into many other physical situations as well. For example, the extension to infinite temperature calculations (see \cite{SM}) seems promising to obtain a more complete understanding of the quantum many-body spectrum. Another avenue of future research could include a study of many-body localized systems after a quantum quench \cite{spec1}, where energy statics are routinely used to classify the mobility edge. At the same time one could apply this tool to obtain a more complete characterization of quantum defects in the fields of quantum computing and counter-diabatic driving.

\textit{Acknowledgements.---} DMK and CK acknowledge support by the Deutsche Forschungsgemeinschaft through the Emmy Noether program (KA 3360/2-1). AJM was supported by the Basic Energy Sciences Division of the U.S. Department of Energy under grant DE-SC0018218. Simulations were performed with computing resources granted by RWTH Aachen University under projects rwth0013 and prep0010.

\bibliography{references,KibbleZurekWavefunction}

\begin{thebibliography}{32}%
\makeatletter
\providecommand \@ifxundefined [1]{%
 \@ifx{#1\undefined}
}%
\providecommand \@ifnum [1]{%
 \ifnum #1\expandafter \@firstoftwo
 \else \expandafter \@secondoftwo
 \fi
}%
\providecommand \@ifx [1]{%
 \ifx #1\expandafter \@firstoftwo
 \else \expandafter \@secondoftwo
 \fi
}%
\providecommand \natexlab [1]{#1}%
\providecommand \enquote  [1]{``#1''}%
\providecommand \bibnamefont  [1]{#1}%
\providecommand \bibfnamefont [1]{#1}%
\providecommand \citenamefont [1]{#1}%
\providecommand \href@noop [0]{\@secondoftwo}%
\providecommand \href [0]{\begingroup \@sanitize@url \@href}%
\providecommand \@href[1]{\@@startlink{#1}\@@href}%
\providecommand \@@href[1]{\endgroup#1\@@endlink}%
\providecommand \@sanitize@url [0]{\catcode `\\12\catcode `\$12\catcode
  `\&12\catcode `\#12\catcode `\^12\catcode `\_12\catcode `\%12\relax}%
\providecommand \@@startlink[1]{}%
\providecommand \@@endlink[0]{}%
\providecommand \url  [0]{\begingroup\@sanitize@url \@url }%
\providecommand \@url [1]{\endgroup\@href {#1}{\urlprefix }}%
\providecommand \urlprefix  [0]{URL }%
\providecommand \Eprint [0]{\href }%
\providecommand \doibase [0]{http://dx.doi.org/}%
\providecommand \selectlanguage [0]{\@gobble}%
\providecommand \bibinfo  [0]{\@secondoftwo}%
\providecommand \bibfield  [0]{\@secondoftwo}%
\providecommand \translation [1]{[#1]}%
\providecommand \BibitemOpen [0]{}%
\providecommand \bibitemStop [0]{}%
\providecommand \bibitemNoStop [0]{.\EOS\space}%
\providecommand \EOS [0]{\spacefactor3000\relax}%
\providecommand \BibitemShut  [1]{\csname bibitem#1\endcsname}%
\let\auto@bib@innerbib\@empty
\bibitem [{\citenamefont {Streltsov}\ \emph {et~al.}(2017)\citenamefont
  {Streltsov}, \citenamefont {Adesso},\ and\ \citenamefont
  {Plenio}}]{Streltsov17}%
  \BibitemOpen
  \bibfield  {author} {\bibinfo {author} {\bibfnamefont {A.}~\bibnamefont
  {Streltsov}}, \bibinfo {author} {\bibfnamefont {G.}~\bibnamefont {Adesso}}, \
  and\ \bibinfo {author} {\bibfnamefont {M.~B.}\ \bibnamefont {Plenio}},\
  }\href {\doibase 10.1103/RevModPhys.89.041003} {\bibfield  {journal}
  {\bibinfo  {journal} {Rev. Mod. Phys.}\ }\textbf {\bibinfo {volume} {89}},\
  \bibinfo {pages} {041003} (\bibinfo {year} {2017})}\BibitemShut {NoStop}%
\bibitem [{\citenamefont {Mankowsky}\ \emph {et~al.}(2016)\citenamefont
  {Mankowsky}, \citenamefont {F{\"o}rst},\ and\ \citenamefont
  {Cavalleri}}]{Mankowsky:Nonequilibrium}%
  \BibitemOpen
  \bibfield  {author} {\bibinfo {author} {\bibfnamefont {R.}~\bibnamefont
  {Mankowsky}}, \bibinfo {author} {\bibfnamefont {M.}~\bibnamefont
  {F{\"o}rst}}, \ and\ \bibinfo {author} {\bibfnamefont {A.}~\bibnamefont
  {Cavalleri}},\ }\href {http://stacks.iop.org/0034-4885/79/i=6/a=064503}
  {\bibfield  {journal} {\bibinfo  {journal} {Rep. on Prog. in Phys.}\ }\textbf
  {\bibinfo {volume} {79}},\ \bibinfo {pages} {064503} (\bibinfo {year}
  {2016})}\BibitemShut {NoStop}%
\bibitem [{\citenamefont {Basov}\ \emph {et~al.}(2017)\citenamefont {Basov},
  \citenamefont {Averitt},\ and\ \citenamefont {Hsieh}}]{Basov:Towards}%
  \BibitemOpen
  \bibfield  {author} {\bibinfo {author} {\bibfnamefont {D.~N.}\ \bibnamefont
  {Basov}}, \bibinfo {author} {\bibfnamefont {R.~D.}\ \bibnamefont {Averitt}},
  \ and\ \bibinfo {author} {\bibfnamefont {D.}~\bibnamefont {Hsieh}},\ }\href
  {\doibase 10.1038/NMAT5017} {\bibfield  {journal} {\bibinfo  {journal}
  {Nature Materials}\ }\textbf {\bibinfo {volume} {16}},\ \bibinfo {pages}
  {1077} (\bibinfo {year} {2017})}\BibitemShut {NoStop}%
\bibitem [{\citenamefont {Tokura}\ \emph {et~al.}(2017)\citenamefont {Tokura},
  \citenamefont {Kawasaki},\ and\ \citenamefont {Nagaosa}}]{Tokura:Emergent}%
  \BibitemOpen
  \bibfield  {author} {\bibinfo {author} {\bibfnamefont {Y.}~\bibnamefont
  {Tokura}}, \bibinfo {author} {\bibfnamefont {M.}~\bibnamefont {Kawasaki}}, \
  and\ \bibinfo {author} {\bibfnamefont {N.}~\bibnamefont {Nagaosa}},\ }\href
  {\doibase 10.1038/nphys4274} {\bibfield  {journal} {\bibinfo  {journal}
  {Nature Phys.}\ }\textbf {\bibinfo {volume} {13}},\ \bibinfo {pages} {1056}
  (\bibinfo {year} {2017})}\BibitemShut {NoStop}%
\bibitem [{\citenamefont {Kibble}(1976)}]{Kibble1976}%
  \BibitemOpen
  \bibfield  {author} {\bibinfo {author} {\bibfnamefont {T.~W.}\ \bibnamefont
  {Kibble}},\ }\href {\doibase 10.1088/0305-4470/9/8/029} {\bibfield  {journal}
  {\bibinfo  {journal} {Journal of Physics A: General Physics}\ }\textbf
  {\bibinfo {volume} {9}},\ \bibinfo {pages} {1387} (\bibinfo {year}
  {1976})}\BibitemShut {NoStop}%
\bibitem [{\citenamefont {Kibble}(1980)}]{Kibble1980}%
  \BibitemOpen
  \bibfield  {author} {\bibinfo {author} {\bibfnamefont {T.~W.}\ \bibnamefont
  {Kibble}},\ }\href {\doibase 10.1016/0370-1573(80)90091-5} {\bibfield
  {journal} {\bibinfo  {journal} {Physics Reports}\ }\textbf {\bibinfo {volume}
  {67}},\ \bibinfo {pages} {183} (\bibinfo {year} {1980})}\BibitemShut
  {NoStop}%
\bibitem [{\citenamefont {Zurek}(1985)}]{Zurek1985}%
  \BibitemOpen
  \bibfield  {author} {\bibinfo {author} {\bibfnamefont {W.~H.}\ \bibnamefont
  {Zurek}},\ }\href {\doibase 10.1038/317505a0} {\bibfield  {journal} {\bibinfo
   {journal} {Nature}\ }\textbf {\bibinfo {volume} {317}},\ \bibinfo {pages}
  {505} (\bibinfo {year} {1985})},\ \Eprint {http://arxiv.org/abs/9607135}
  {9607135} \BibitemShut {NoStop}%
\bibitem [{\citenamefont {Dutta}\ \emph {et~al.}(2015)\citenamefont {Dutta},
  \citenamefont {Aeppli}, \citenamefont {Chakrabarti}, \citenamefont
  {Divakaran}, \citenamefont {Rosenbaum},\ and\ \citenamefont
  {Sen}}]{Dutta2010}%
  \BibitemOpen
  \bibfield  {author} {\bibinfo {author} {\bibfnamefont {A.}~\bibnamefont
  {Dutta}}, \bibinfo {author} {\bibfnamefont {G.}~\bibnamefont {Aeppli}},
  \bibinfo {author} {\bibfnamefont {B.~K.}\ \bibnamefont {Chakrabarti}},
  \bibinfo {author} {\bibfnamefont {U.}~\bibnamefont {Divakaran}}, \bibinfo
  {author} {\bibfnamefont {T.~F.}\ \bibnamefont {Rosenbaum}}, \ and\ \bibinfo
  {author} {\bibfnamefont {D.}~\bibnamefont {Sen}},\ }\href {\doibase
  10.1017/CBO9781107706057} {\emph {\bibinfo {title} {{Quantum Phase
  Transitions in Transverse Field Spin Models}}}}\ (\bibinfo {year} {2015})\
  \Eprint {http://arxiv.org/abs/1012.0653} {arXiv:1012.0653} \BibitemShut
  {NoStop}%
\bibitem [{\citenamefont {Dziarmaga}(2010)}]{Dziarmaga2010}%
  \BibitemOpen
  \bibfield  {author} {\bibinfo {author} {\bibfnamefont {J.}~\bibnamefont
  {Dziarmaga}},\ }\href {\doibase 10.1080/00018732.2010.514702} {\bibfield
  {journal} {\bibinfo  {journal} {Advances in Physics}\ }\textbf {\bibinfo
  {volume} {59}},\ \bibinfo {pages} {1063} (\bibinfo {year} {2010})},\ \Eprint
  {http://arxiv.org/abs/https://doi.org/10.1080/00018732.2010.514702}
  {https://doi.org/10.1080/00018732.2010.514702} \BibitemShut {NoStop}%
\bibitem [{\citenamefont {del Campo}\ and\ \citenamefont
  {Zurek}(2013)}]{DelCampo2014}%
  \BibitemOpen
  \bibfield  {author} {\bibinfo {author} {\bibfnamefont {A.}~\bibnamefont {del
  Campo}}\ and\ \bibinfo {author} {\bibfnamefont {W.~H.}\ \bibnamefont
  {Zurek}},\ }\href {\doibase 10.1142/S0217751X1430018X} {\bibfield  {journal}
  {\bibinfo  {journal} {International Journal of Modern Physics A}\ }\textbf
  {\bibinfo {volume} {29}},\ \bibinfo {pages} {1430018} (\bibinfo {year}
  {2013})},\ \Eprint {http://arxiv.org/abs/1310.1600} {arXiv:1310.1600}
  \BibitemShut {NoStop}%
\bibitem [{\citenamefont {Quan}\ and\ \citenamefont {Zurek}(2010)}]{Quan2010}%
  \BibitemOpen
  \bibfield  {author} {\bibinfo {author} {\bibfnamefont {H.~T.}\ \bibnamefont
  {Quan}}\ and\ \bibinfo {author} {\bibfnamefont {W.~H.}\ \bibnamefont
  {Zurek}},\ }\href {\doibase 10.1088/1367-2630/12/9/093025} {\bibfield
  {journal} {\bibinfo  {journal} {New Journal of Physics}\ }\textbf {\bibinfo
  {volume} {12}},\ \bibinfo {pages} {093025} (\bibinfo {year} {2010})},\
  \Eprint {http://arxiv.org/abs/1007.3294} {arXiv:1007.3294} \BibitemShut
  {NoStop}%
\bibitem [{\citenamefont {Peres}(1984)}]{Peres1984}%
  \BibitemOpen
  \bibfield  {author} {\bibinfo {author} {\bibfnamefont {A.}~\bibnamefont
  {Peres}},\ }\href {\doibase 10.1103/PhysRevA.30.1610} {\bibfield  {journal}
  {\bibinfo  {journal} {Physical Review A}\ }\textbf {\bibinfo {volume} {30}},\
  \bibinfo {pages} {1610} (\bibinfo {year} {1984})}\BibitemShut {NoStop}%
\bibitem [{\citenamefont {Pastawski}\ \emph {et~al.}(1995)\citenamefont
  {Pastawski}, \citenamefont {Levstein},\ and\ \citenamefont
  {Usaj}}]{Pastawski1995}%
  \BibitemOpen
  \bibfield  {author} {\bibinfo {author} {\bibfnamefont {H.~M.}\ \bibnamefont
  {Pastawski}}, \bibinfo {author} {\bibfnamefont {P.~R.}\ \bibnamefont
  {Levstein}}, \ and\ \bibinfo {author} {\bibfnamefont {G.}~\bibnamefont
  {Usaj}},\ }\href {\doibase 10.1103/PhysRevLett.75.4310} {\bibfield  {journal}
  {\bibinfo  {journal} {Physical Review Letters}\ }\textbf {\bibinfo {volume}
  {75}},\ \bibinfo {pages} {4310} (\bibinfo {year} {1995})},\ \Eprint
  {http://arxiv.org/abs/9604019} {arXiv:9604019 [cond-mat]} \BibitemShut
  {NoStop}%
\bibitem [{\citenamefont {Silva}(2008)}]{Silva2008}%
  \BibitemOpen
  \bibfield  {author} {\bibinfo {author} {\bibfnamefont {A.}~\bibnamefont
  {Silva}},\ }\href {\doibase 10.1103/PhysRevLett.101.120603} {\bibfield
  {journal} {\bibinfo  {journal} {Physical Review Letters}\ }\textbf {\bibinfo
  {volume} {101}},\ \bibinfo {pages} {120603} (\bibinfo {year} {2008})},\
  \Eprint {http://arxiv.org/abs/0806.4301} {arXiv:0806.4301} \BibitemShut
  {NoStop}%
\bibitem [{\citenamefont {Chenu}\ \emph {et~al.}(2019)\citenamefont {Chenu},
  \citenamefont {Molina-Vilaplana},\ and\ \citenamefont {del
  Campo}}]{Chenu2017}%
  \BibitemOpen
  \bibfield  {author} {\bibinfo {author} {\bibfnamefont {A.}~\bibnamefont
  {Chenu}}, \bibinfo {author} {\bibfnamefont {J.}~\bibnamefont
  {Molina-Vilaplana}}, \ and\ \bibinfo {author} {\bibfnamefont
  {A.}~\bibnamefont {del Campo}},\ }\href {http://arxiv.org/abs/1711.01277
  http://arxiv.org/abs/1804.09188} {\bibfield  {journal} {\bibinfo  {journal}
  {Quantum}\ }\textbf {\bibinfo {volume} {3}},\ \bibinfo {pages} {127}
  (\bibinfo {year} {2019})},\ \Eprint {http://arxiv.org/abs/1804.09188}
  {arXiv:1804.09188} \BibitemShut {NoStop}%
\bibitem [{\citenamefont {Pandey}\ \emph {et~al.}(2018)\citenamefont {Pandey},
  \citenamefont {Plekhanov},\ and\ \citenamefont {Roux}}]{spec1}%
  \BibitemOpen
  \bibfield  {author} {\bibinfo {author} {\bibfnamefont {B.}~\bibnamefont
  {Pandey}}, \bibinfo {author} {\bibfnamefont {K.}~\bibnamefont {Plekhanov}}, \
  and\ \bibinfo {author} {\bibfnamefont {G.}~\bibnamefont {Roux}},\ }\href
  {\doibase 10.1103/PhysRevA.98.050103} {\bibfield  {journal} {\bibinfo
  {journal} {Phys. Rev. A}\ }\textbf {\bibinfo {volume} {98}},\ \bibinfo
  {pages} {050103} (\bibinfo {year} {2018})}\BibitemShut {NoStop}%
\bibitem [{\citenamefont {White}(1992)}]{White1992}%
  \BibitemOpen
  \bibfield  {author} {\bibinfo {author} {\bibfnamefont {S.~R.}\ \bibnamefont
  {White}},\ }\href {\doibase 10.1103/PhysRevLett.69.2863} {\bibfield
  {journal} {\bibinfo  {journal} {Physical Review Letters}\ }\textbf {\bibinfo
  {volume} {69}},\ \bibinfo {pages} {2863} (\bibinfo {year}
  {1992})}\BibitemShut {NoStop}%
\bibitem [{\citenamefont {Vidal}(2007)}]{Vidal2007a}%
  \BibitemOpen
  \bibfield  {author} {\bibinfo {author} {\bibfnamefont {G.}~\bibnamefont
  {Vidal}},\ }\href {\doibase 10.1103/PhysRevLett.98.070201} {\bibfield
  {journal} {\bibinfo  {journal} {Phys Rev Lett}\ }\textbf {\bibinfo {volume}
  {98}},\ \bibinfo {pages} {70201} (\bibinfo {year} {2007})}\BibitemShut
  {NoStop}%
\bibitem [{\citenamefont {Schollw{\"{o}}ck}(2011)}]{Schollwock2011a}%
  \BibitemOpen
  \bibfield  {author} {\bibinfo {author} {\bibfnamefont {U.}~\bibnamefont
  {Schollw{\"{o}}ck}},\ }\href {\doibase 10.1016/j.aop.2010.09.012} {\bibfield
  {journal} {\bibinfo  {journal} {Annals of Physics}\ }\textbf {\bibinfo
  {volume} {326}},\ \bibinfo {pages} {96} (\bibinfo {year} {2011})},\ \Eprint
  {http://arxiv.org/abs/1008.3477} {arXiv:1008.3477} \BibitemShut {NoStop}%
\bibitem [{\citenamefont {Kennes}\ and\ \citenamefont
  {Karrasch}(2016)}]{Kennes2016a}%
  \BibitemOpen
  \bibfield  {author} {\bibinfo {author} {\bibfnamefont {D.~M.}\ \bibnamefont
  {Kennes}}\ and\ \bibinfo {author} {\bibfnamefont {C.}~\bibnamefont
  {Karrasch}},\ }\href {\doibase 10.1016/j.cpc.2015.10.019} {\bibfield
  {journal} {\bibinfo  {journal} {Computer Physics Communications}\ }\textbf
  {\bibinfo {volume} {200}},\ \bibinfo {pages} {37} (\bibinfo {year} {2016})},\
  \Eprint {http://arxiv.org/abs/1404.3704} {arXiv:1404.3704} \BibitemShut
  {NoStop}%
\bibitem [{SM()}]{SM}%
  \BibitemOpen
  \href@noop {} {}\bibinfo {note} {See Supplemental Material at [URL will be
  inserted by publisher] for details of the ramp, the algorithm and further
  analysis.}\BibitemShut {Stop}%
\bibitem [{\citenamefont {Lieb}\ \emph {et~al.}(1961)\citenamefont {Lieb},
  \citenamefont {Schultz},\ and\ \citenamefont {Mattis}}]{Lieb1961}%
  \BibitemOpen
  \bibfield  {author} {\bibinfo {author} {\bibfnamefont {E.}~\bibnamefont
  {Lieb}}, \bibinfo {author} {\bibfnamefont {T.}~\bibnamefont {Schultz}}, \
  and\ \bibinfo {author} {\bibfnamefont {D.}~\bibnamefont {Mattis}},\ }\href
  {\doibase 10.1016/0003-4916(61)90115-4} {\bibfield  {journal} {\bibinfo
  {journal} {Annals of Physics}\ }\textbf {\bibinfo {volume} {16}},\ \bibinfo
  {pages} {407} (\bibinfo {year} {1961})}\BibitemShut {NoStop}%
\bibitem [{\citenamefont {Pfeuty}(1970)}]{Pfeuty1970}%
  \BibitemOpen
  \bibfield  {author} {\bibinfo {author} {\bibfnamefont {P.}~\bibnamefont
  {Pfeuty}},\ }\href {\doibase 10.1016/0003-4916(70)90270-8} {\bibfield
  {journal} {\bibinfo  {journal} {Annals of Physics}\ }\textbf {\bibinfo
  {volume} {57}},\ \bibinfo {pages} {79} (\bibinfo {year} {1970})}\BibitemShut
  {NoStop}%
\bibitem [{\citenamefont {He}\ and\ \citenamefont {Guo}(2017)}]{He2017}%
  \BibitemOpen
  \bibfield  {author} {\bibinfo {author} {\bibfnamefont {Y.}~\bibnamefont
  {He}}\ and\ \bibinfo {author} {\bibfnamefont {H.}~\bibnamefont {Guo}},\
  }\href {\doibase 10.1088/1742-5468/aa85b0} {\bibfield  {journal} {\bibinfo
  {journal} {Journal of Statistical Mechanics: Theory and Experiment}\ }\textbf
  {\bibinfo {volume} {2017}},\ \bibinfo {pages} {093101} (\bibinfo {year}
  {2017})},\ \Eprint {http://arxiv.org/abs/1707.02400} {arXiv:1707.02400}
  \BibitemShut {NoStop}%
\bibitem [{\citenamefont {Gell-Mann}\ and\ \citenamefont
  {Low}(1951)}]{Gell-Mann1951}%
  \BibitemOpen
  \bibfield  {author} {\bibinfo {author} {\bibfnamefont {M.}~\bibnamefont
  {Gell-Mann}}\ and\ \bibinfo {author} {\bibfnamefont {F.}~\bibnamefont
  {Low}},\ }\href {\doibase 10.1103/PhysRev.84.350} {\bibfield  {journal}
  {\bibinfo  {journal} {Physical Review}\ }\textbf {\bibinfo {volume} {84}},\
  \bibinfo {pages} {350} (\bibinfo {year} {1951})},\ \Eprint
  {http://arxiv.org/abs/9907483} {arXiv:9907483 [hep-ph]} \BibitemShut
  {NoStop}%
\bibitem [{\citenamefont {Brouder}\ \emph {et~al.}(2008)\citenamefont
  {Brouder}, \citenamefont {Stoltz},\ and\ \citenamefont
  {Panati}}]{Brouder2008}%
  \BibitemOpen
  \bibfield  {author} {\bibinfo {author} {\bibfnamefont {C.}~\bibnamefont
  {Brouder}}, \bibinfo {author} {\bibfnamefont {G.}~\bibnamefont {Stoltz}}, \
  and\ \bibinfo {author} {\bibfnamefont {G.}~\bibnamefont {Panati}},\ }\href
  {\doibase 10.1103/PhysRevA.78.042102} {\bibfield  {journal} {\bibinfo
  {journal} {Physical Review A - Atomic, Molecular, and Optical Physics}\
  }\textbf {\bibinfo {volume} {78}},\ \bibinfo {pages} {042102} (\bibinfo
  {year} {2008})},\ \Eprint {http://arxiv.org/abs/0807.4218} {arXiv:0807.4218}
  \BibitemShut {NoStop}%
\bibitem [{\citenamefont {Bachmann}\ \emph {et~al.}(2017)\citenamefont
  {Bachmann}, \citenamefont {De~Roeck},\ and\ \citenamefont {Fraas}}]{adia1}%
  \BibitemOpen
  \bibfield  {author} {\bibinfo {author} {\bibfnamefont {S.}~\bibnamefont
  {Bachmann}}, \bibinfo {author} {\bibfnamefont {W.}~\bibnamefont {De~Roeck}},
  \ and\ \bibinfo {author} {\bibfnamefont {M.}~\bibnamefont {Fraas}},\ }\href
  {\doibase 10.1103/PhysRevLett.119.060201} {\bibfield  {journal} {\bibinfo
  {journal} {Phys. Rev. Lett.}\ }\textbf {\bibinfo {volume} {119}},\ \bibinfo
  {pages} {060201} (\bibinfo {year} {2017})}\BibitemShut {NoStop}%
\bibitem [{\citenamefont {Kennes}(2017)}]{adia2}%
  \BibitemOpen
  \bibfield  {author} {\bibinfo {author} {\bibfnamefont {D.~M.}\ \bibnamefont
  {Kennes}},\ }\href {\doibase 10.1103/PhysRevB.96.024302} {\bibfield
  {journal} {\bibinfo  {journal} {Phys. Rev. B}\ }\textbf {\bibinfo {volume}
  {96}},\ \bibinfo {pages} {024302} (\bibinfo {year} {2017})}\BibitemShut
  {NoStop}%
\bibitem [{\citenamefont {Lychkovskiy}\ \emph {et~al.}(2017)\citenamefont
  {Lychkovskiy}, \citenamefont {Gamayun},\ and\ \citenamefont
  {Cheianov}}]{adia3}%
  \BibitemOpen
  \bibfield  {author} {\bibinfo {author} {\bibfnamefont {O.}~\bibnamefont
  {Lychkovskiy}}, \bibinfo {author} {\bibfnamefont {O.}~\bibnamefont
  {Gamayun}}, \ and\ \bibinfo {author} {\bibfnamefont {V.}~\bibnamefont
  {Cheianov}},\ }\href {\doibase 10.1103/PhysRevLett.119.200401} {\bibfield
  {journal} {\bibinfo  {journal} {Phys. Rev. Lett.}\ }\textbf {\bibinfo
  {volume} {119}},\ \bibinfo {pages} {200401} (\bibinfo {year}
  {2017})}\BibitemShut {NoStop}%
\bibitem [{\citenamefont {Dziarmaga}(2005)}]{Dziarmaga05}%
  \BibitemOpen
  \bibfield  {author} {\bibinfo {author} {\bibfnamefont {J.}~\bibnamefont
  {Dziarmaga}},\ }\href {\doibase 10.1103/PhysRevLett.95.245701} {\bibfield
  {journal} {\bibinfo  {journal} {Phys. Rev. Lett.}\ }\textbf {\bibinfo
  {volume} {95}},\ \bibinfo {pages} {245701} (\bibinfo {year}
  {2005})}\BibitemShut {NoStop}%
\bibitem [{\citenamefont {Kennes}\ \emph {et~al.}(2018)\citenamefont {Kennes},
  \citenamefont {de~la Torre}, \citenamefont {Ron}, \citenamefont {Hsieh},\
  and\ \citenamefont {Millis}}]{Kennes2018}%
  \BibitemOpen
  \bibfield  {author} {\bibinfo {author} {\bibfnamefont {D.~M.}\ \bibnamefont
  {Kennes}}, \bibinfo {author} {\bibfnamefont {A.}~\bibnamefont {de~la Torre}},
  \bibinfo {author} {\bibfnamefont {A.}~\bibnamefont {Ron}}, \bibinfo {author}
  {\bibfnamefont {D.}~\bibnamefont {Hsieh}}, \ and\ \bibinfo {author}
  {\bibfnamefont {A.~J.}\ \bibnamefont {Millis}},\ }\href {\doibase
  10.1103/PhysRevLett.120.127601} {\bibfield  {journal} {\bibinfo  {journal}
  {Phys. Rev. Lett.}\ }\textbf {\bibinfo {volume} {120}},\ \bibinfo {pages}
  {127601} (\bibinfo {year} {2018})}\BibitemShut {NoStop}%
\bibitem [{\citenamefont {Bernien}\ \emph {et~al.}(2017)\citenamefont
  {Bernien}, \citenamefont {Schwartz}, \citenamefont {Keesling}, \citenamefont
  {Levine}, \citenamefont {Omran}, \citenamefont {Pichler}, \citenamefont
  {Choi}, \citenamefont {Zibrov}, \citenamefont {Endres}, \citenamefont
  {Greiner}, \citenamefont {Vuletic},\ and\ \citenamefont
  {Lukin}}]{Bernien2017}%
  \BibitemOpen
  \bibfield  {author} {\bibinfo {author} {\bibfnamefont {H.}~\bibnamefont
  {Bernien}}, \bibinfo {author} {\bibfnamefont {S.}~\bibnamefont {Schwartz}},
  \bibinfo {author} {\bibfnamefont {A.}~\bibnamefont {Keesling}}, \bibinfo
  {author} {\bibfnamefont {H.}~\bibnamefont {Levine}}, \bibinfo {author}
  {\bibfnamefont {A.}~\bibnamefont {Omran}}, \bibinfo {author} {\bibfnamefont
  {H.}~\bibnamefont {Pichler}}, \bibinfo {author} {\bibfnamefont
  {S.}~\bibnamefont {Choi}}, \bibinfo {author} {\bibfnamefont {A.~S.}\
  \bibnamefont {Zibrov}}, \bibinfo {author} {\bibfnamefont {M.}~\bibnamefont
  {Endres}}, \bibinfo {author} {\bibfnamefont {M.}~\bibnamefont {Greiner}},
  \bibinfo {author} {\bibfnamefont {V.}~\bibnamefont {Vuletic}}, \ and\
  \bibinfo {author} {\bibfnamefont {M.~D.}\ \bibnamefont {Lukin}},\ }\href
  {\doibase 10.1038/nature24622} {\bibfield  {journal} {\bibinfo  {journal}
  {Nature}\ }\textbf {\bibinfo {volume} {551}},\ \bibinfo {pages} {579}
  (\bibinfo {year} {2017})},\ \Eprint {http://arxiv.org/abs/1707.04344}
  {arXiv:1707.04344} \BibitemShut {NoStop}%
\end{thebibliography}%

\end{document}